\documentclass[pss]{wiley2sp} 
\usepackage{amsmath}

\begin{document}

\title{Reversing the Landauer's erasure: single-electron Maxwell's demon operating at the limit of thermodynamic efficiency}

\titlerunning{Reversible Maxwell's demon}

\author{%
  Dmitri V. Averin\textsuperscript{\Ast,\textsf{\bfseries 1}} and
  Jukka P. Pekola\textsuperscript{\textsf{\bfseries 2}}}

\authorrunning{Dmitri Averin et al.}

\mail{e-mail
  \textsf{dmitri.averin@stonybrook.edu}}

\institute{%
\textsuperscript{1}\, Department of Physics and Astronomy, Stony Brook
University, SUNY, Stony Brook, NY 11794-3800 \\
\textsuperscript{2}\, Low Temperature Laboratory, Department of Applied Physics, Aalto University School of Science, P.O. Box 13500, 00076 Aalto, Finland}

\received{XXXX, revised XXXX, accepted XXXX} 
\published{XXXX} 


\abstract{%
%
%
%
\abstcol{%
According to Landauer's principle, erasure of information is the only part of a computation process that unavoidably involves energy dissipation. If done reversibly, such an erasure generates the minimal heat of $k_BT\ln 2$ per erased bit of information. The goal of this
work is to discuss the actual reversal of the optimal erasure which can serve as the basis for the Maxwell's demon operating with ultimate thermodynamic efficiency as dictated by the second law of thermodynamics. The demon extracts $k_BT\ln 2$ of heat from an equilibrium reservoir at temperature $T$ per one bit of information obtained about the measured system used by the demon. We have analyzed this Maxwell's demon in the situation when it uses a general quantum system with a discrete spectrum of energy levels as its working body.
  }{%
In the case of the effectively two-level system, which has been realized experimentally based on tunneling of individual electron in a single-electron box \cite{we1}, we also studied and minimized corrections to the ideal reversible operation of the demon. These corrections include, in particular, the non-adiabatic terms which are described by a version of the classical fluctuation-dissipation theorem. The overall reversibility of the Maxwell's demon requires,
beside the reversibility of the intrinsic working body dynamics, the reversibility of the measurement and feedback processes. The single-electron demon can, in principle, be made fully reversible by developing a thermodynamically reversible single-electron charge detector for measurements of the individual charge states of the single-electron box.
 }}

%
%

\maketitle   

\section{Introduction}

Thermodynamics of nanostructures, in which quantum mechanics and statistical fluctuations play an important role, has recently attracted considerable attention, see, e.g., the reviews \cite{bustamante,campisi,seifert,pekola,millen} and references therein. One of the most interesting issues in this field is realization of Maxwell's demon \cite{toyabe,we1,koski,QMD}, and more generally, understanding the role of information in thermodynamic processes \cite{pt}. Development of thermodynamics of information is also motivated by practical attempt to demonstrate thermodynamically reversible computation -- see, e.g., \cite{rev}.

The foundation for the physics of information was provided by Rolf Landauer in the context of thermodynamics of computation. He demonstrated \cite{landauer} that erasure of information is the only part of a computation process that unavoidably involves energy dissipation. Erasure of one bit of information consists in bringing a two-state system initially in the most uncertain configuration (both states occupied with probability $1/2$) to the pre-determined  configuration, when the system is in one of its states with probability 1. This process leads to generation of heat in the reservoir at temperature $T$, with the minimum of generated heat, $k_BT\ln2$, achieved if the erasure is performed in the optimal, i.e. reversible, way, as was recently confirmed experimentally \cite{exp1,exp2,exp3}.

It has been understood for quite some time (see, e.g., \cite{bennett}) that the principle of information erasure, and not the energy dissipation in the measurement process as believed previously \cite{brillouin}, provides the resolution of the Maxwell's demon paradox, reconciling operation of this device with the second law of thermodynamics. When the information collected by the demon in the process of its operation is erased in order to return the system to the initial state, this information erasure process dissipates back into the reservoir in the form of heat the same amount of free energy as extracted by the demon.
The purpose of this work is to make this relation between the reversible Landauer's information erasure and Maxwell's demon even more close, by discussing the actual reversal of the optimal erasure process, which, if complemented with a measurement and feedback, produces the Maxwell’s demon operating with the highest possible thermodynamic efficiency. The demon extracts $k_BT\ln 2$ of heat from an equilibrium reservoir at temperature $T$ at the cost of creating one bit of information about the state of the system which serves as its working body. Operation of such a reversible Maxwell’s demon was demonstrated recently \cite{we1} using an individual electron charge on the single-electron box as the two-state working substance. This demonstration gives an example of a general use of the single-charge structures \cite{al1} which provide a convenient setting for studying various aspects of non-equilibrium nanoscale thermodynamics \cite{we3,kung,saira}.

The main general view on the physics of information that emerged from the original theoretical studies of reversible computation and developed recently in details in the context of the nanoscale thermodynamics, is that the information can be viewed appropriately as the entropy of a computing device \cite{pt}. Although a computing device is quite different from a generic statistical system in that the main degrees of freedom in it are well-controlled and not fluctuating in time, intrinsic randomness of bit strings in the computation process makes it necessary to view them as carrying entropy in the same way as the fluctuating degrees of freedom of a statistical system. This view of information as entropy unites in a simple way understanding of many different phenomena in information-assisted thermodynamics, and has many implications for the physics of computation. One example is the basic notion that the logically irreversible computation that does not conserve information, discarding a part of it during the computational steps, cannot be achieved without the dissipation of at least $k_BT\ln 2$ of energy per discarded bit, despite claims to the contrary in the literature \cite{irrev}. By the second law of thermodynamics, if the information is entropy, reduction of information in the computing device should be inevitably accompanied by the entropy increase in the surrounding environment, the process that can not be achieved without energy dissipation. Understanding of information as entropy also poses new interesting questions about the nature of randomness and thermalization which require further studies.

In this work, we illustrate the connection between the entropy and information by considering the reversible Maxwell's demon based on a general multi-level quantum system. The results here generalize
to the multi-level situation the discussion in Refs.~\cite{we1,we3} of the Maxwell's demon operation and thermodynamics properties of the two-state state systems. The description of the Maxwell's demon is also extended by developing the minimization scheme of the heat dissipated in its operation, which can be essential for reaching the ultimate thermodynamic limit of the Maxwell's demon efficiency.

\section{Thermodynamics of the master equation}

\subsection{First and second laws}

\begin{figure}[t]
\centering
\includegraphics[bb=0 15 520 440, clip=true, scale=0.33]{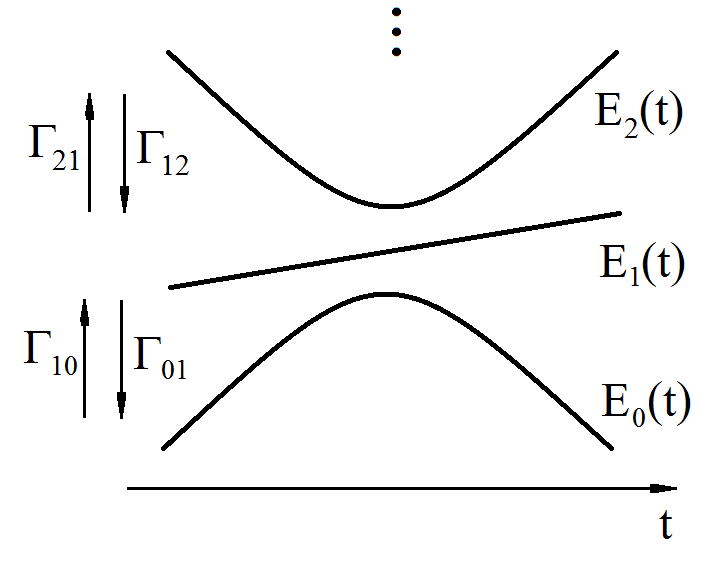}
\caption{Schematics of a general system with discrete energy spectrum with energies
$E_0(t), E_1(t), ...$ adiabatically changing in time. Occupation probabilities of the
energy states of the system evolve according to the master equation \protect (\ref{e1})
due to transitions with rates $\Gamma_{mn}$ induced by a thermal reservoir in equilibrium
at temperature $T$. }
\label{diag} \end{figure}

For completeness, we start our quantitative discussion with a theoretical description of thermodynamic properties of the systems, evolution of which is governed by a rate equation -- see, e.g., \cite{strasberg,we1}. Single-charge tunneling in tunnel junction structures \cite{al1}, e.g., in a single-Cooper-pair \cite{mb} or single-electron box \cite{laf}, provides one of the better-known and precisely-controlled experimental examples of this case. The system we consider is assumed to have a discrete set of energy states $|n\rangle$ with energies $E_n$ controlled by some time-dependent parameters making the energies $E_n(t)$ also time-dependent, as illustrated in Fig.~(\ref{diag}). If the system is in the state $n$, an external source inducing a small variation of the parameter $q$, slow on the time scale set by the energies $E_n$ themselves so as not to induce any transitions out of the state $n$, does work
$dW= dE_n$ on the system. The average work done is then $\langle dW \rangle=\sum_n p_n dE_n$, where $p_n$ is the probability for the system to be in the state $n$, and the notation $\langle ... \rangle= \sum_n p_n ...$ will be used for all quantities. As is typical for experiments with many nanoscale systems, an equilibrium reservoir at temperature $T$ is assumed to interact weakly with the system, inducing stochastic transitions among the energy states $|n\rangle$. The evolution of the probabilities $p_n$ is governed then by the usual rate equation:
\begin{equation}
\dot{p}_n =\sum_{m} (\Gamma_{nm} p_m -\Gamma_{mn} p_n) \, ,
\label{e1} \end{equation}
where the rate $\Gamma_{mn}$ describes the transition from state $|n\rangle$ to state $|m\rangle$ -- see Fig.~(\ref{diag}). In the absence of degeneracies in the energy spectrum of the system, dynamics of a general quantum system can be described with the rate equation for sufficiently weak coupling to the reservoir and sufficiently slow evolution of the system's energies. When a degeneracy is present, validity of the rate equation requires in addition that the operator coupling the system to the reservoir is diagonal in the basis of degenerate energy eigenstates, condition that is satisfied in the examples of the single-electron systems considered in this work. If some of these conditions is not satisfied, off-diagonal elements of the density matrix of the system starts playing role in its dynamics -- see, e.g., \cite{strasberg}. For the reservoir in equilibrium, the transition rates  $\Gamma$ satisfy the detailed balance condition linking the rates of the direct and reverse transitions:
\begin{equation}
\Gamma_{mn}/\Gamma_{nm}= e^{(E_n-E_m)/k_BT}\, .
\label{e2} \end{equation}

Each transition from state $|n\rangle$ to state $|m\rangle$ releases the energy $E_n-E_m$ from the system into the reservoir. Since this process is not associated with any variation of the macroscopic parameters of the system, it represents the heat flow. This means that quantitatively, the contribution of the transitions of this type to the heat $dQ$ flowing from the system into the reservoir during a small time interval $dt$ can be written as $dQ=(E_n-E_m)\Gamma_{mn}dt$, where $\Gamma_{mn}dt$ is the probability that the transition $|n\rangle \rightarrow |m\rangle$ happens during the small interval $dt$. Taking into account that for a given state $n$ transition can take the system into any state $m$, we can write the total heat flowing from the system into the reservoir, when the system is in the state $n$, as
\begin{equation}
dQ=\sum_m(E_n-E_m)\Gamma_{mn}dt\, .
\label{heat} \end{equation}
These relations for heat and work, together with the master equation (\ref{e1}), give the first law of thermodynamics satisfied by the evolution of the system. Indeed, taking the increment of the internal energy $U=\sum_n E_n p_n$ and using the master equation to transform the increment of probability $p_n$, we have
\begin{eqnarray}
dU = \sum_n (dE_n p_n +E_n dp_n) \;\; \nonumber \\
 =\langle dW \rangle + \sum_{n,m} E_n(\Gamma_{nm} p_m -\Gamma_{mn} p_n) dt\;\; \label{first} \\
=\langle dW \rangle + \sum_{n,m} (E_m-E_n)\Gamma_{mn} p_n dt =  \langle dW \rangle-\langle dQ \rangle\, . \nonumber  \end{eqnarray}

To formulate the second law of thermodynamics for the system considered here, one needs to define its entropy. The entropy $S$ of the total structure, including the system of interest and the reservoir, consists of the two parts, Boltzmann's entropy $S_{sys}$ of the probability distribution $\{ p_n\}$ of the system over the states $|n\rangle$:
\begin{equation}
S_{sys} = - k_B \sum_{n} p_n \ln p_n  \, ,
\label{e3} \end{equation}
and the entropy $S_{env}$ of the environment. While $S_{env}$ is not known without specifying explicitly the model of the environment, condition that the environment is in equilibrium at temperature $T$ provides sufficient information about it. Under these conditions,
the change of entropy $S_{env}$ due to each transition $|n\rangle \rightarrow |m\rangle$ can be related directly to the heat transferred into the environment as a result of this transition. The average over all possible transitions then gives:
\begin{equation}
dS_{env} = \frac{\langle dQ \rangle}{T}=\frac{1}{T} \sum_{n,m} (E_n-E_m)
\Gamma_{mn} p_n dt  \, . \;
\label{e4} \end{equation}
Combining these two parts of the total entropy $S$, one can determine its dynamics. The time evolution of the system entropy $S_{sys}$ follows directly from the master equation (\ref{e1}) and the normalization condition $\sum_n p_n=1$:
\begin{eqnarray}
\frac{1}{k_B} \frac{\partial S_{sys}}{\partial t} = - \sum_{n} \dot{p}_n (1+\ln p_n)
\nonumber \\ = \sum_{n,m} (\Gamma_{mn} p_n - \Gamma_{nm} p_m) \ln p_n  \label{ev1} \\
= \frac{1}{2} \sum_{n,m} (\Gamma_{mn} p_n - \Gamma_{nm} p_m) (\ln p_n - \ln p_m) \, .
\nonumber \end{eqnarray}
Evolution equation for the environment entropy $S_{env}$ which describes the evolution of $S_{env}$ resulting from the system dynamics, follows directly from Eq.~(\ref{e4}), and
can be transformed as follows:
\begin{eqnarray}
\frac{\partial S_{env}}{\partial t} = \frac{1}{T} \sum_{n,m} (E_n-E_m) \Gamma_{mn} p_n
\nonumber \\ = \frac{1}{2T} \sum_{n,m} (E_n-E_m) (\Gamma_{mn} p_n - \Gamma_{nm} p_m)  \, .
\label{ev2} \end{eqnarray}

Adding Eqs.~(\ref{ev1}) and (\ref{ev2}), we get the equation that governs the time evolution of the total entropy $S$:
\begin{equation}
\frac{\partial S}{\partial t} = \frac{1}{2} \sum_{n,m} [k_B \ln \frac{p_n}{p_m} +\frac{E_n-E_m}{T}] (\Gamma_{mn}p_n -\Gamma_{nm}p_m  ) \, .
\label{ev} \end{equation}
This equation can be simplified further using the detailed-balance condition (\ref{e2}) in the form $(E_n-E_m)/k_B T =\ln (\Gamma_{mn}/\Gamma_{nm})$ to transform the last term in the brackets. In this way, Eq.~(\ref{ev}) gives the rate of the total entropy production in the course of the time evolution described by the master equation (\ref{e1}):
\begin{equation}
\frac{\partial S}{\partial t} = \frac{k_B}{2} \sum_{n,m} \ln
\frac{\Gamma_{mn}p_n}{ \Gamma_{nm}p_m} (\Gamma_{mn}p_n -\Gamma_{nm}p_m )  \, .
\label{second} \end{equation}
This equation shows explicitly that $\partial S/\partial t\geq 0$, and provides the expression of the second law of thermodynamics for the evolution governed by the master equation.

\vspace{1ex}

\subsection{Adiabatic evolution and dissipated heat and work}

Adiabatic evolution, characterized by the vanishingly small rate $\eta$ of variation of the energies of the system, $\eta \sim \dot{E}_n/E_n \rightarrow 0$, conserves the total entropy $S$, and is of the main interest for all reversible processes, in particular for the subsequent discussion of the reversible Maxwell's demon. In this regime, the rate of the variation of probabilities $p_n$ is also small, $\dot{p}_n \propto \eta \rightarrow 0$ and they in the main approximation are given by the stationary solution of the master equation (\ref{e1}). In the typical case when the energy spectrum $\{ E_n\}$ of the system includes the state with the lowest energy, such a stationary state corresponds to equilibrium, i.e., all the probability fluxes in the master equation vanish: $\Gamma_{mn}p_n =p_m \Gamma_{nm}$. For an equilibrium reservoir characterized by Eq.~(\ref{e2}), this relation implies that the probabilities $p_n$ indeed maintain the equilibrium form, $p_n/p_m = e^{[E_n(t)-E_m(t)]/T}$ throughout the evolution.  Equation (\ref{second}) shows then that the total entropy production vanishes in this case, $\partial S/\partial t =0$, i.e., $\partial S_{sys}/\partial t = -\partial S_{env}/\partial t$, implying that the changes $\Delta S_{sys}$ of the entropy of the system are directly compensated by the average heat flow $\langle Q\rangle^{(rev)}$ to (or from) the reservoir:
\begin{equation}
\langle Q\rangle^{(rev)} = -T \Delta S_{sys} \, .
\label{eq} \end{equation}
Since the total entropy is conserved, this transfer of heat into the reservoir represents reversible process which can be inverted, returning the system to the initial state and removing the transferred heat from the reservoir.

While the relation (\ref{eq}) holds in the limit of vanishing rate of variations in the system, $\eta \rightarrow 0$, finite $\eta$ makes the process irreversible, and generates additional ``dissipated'' heat in the reservoir. To find such heat in a general situation, it is convenient to write the master equation (\ref{e1}) in the matrix form, $\dot{p} = \gamma p$, where $p$ is the vector column of the probabilities $p_n$, and $\gamma$ is the matrix of the transition rates with the matrix elements given by the two relations: $\gamma_{nm}= \Gamma_{nm}$, for $n\neq m$, and $\gamma_{nn}=-\sum_{m\neq n} \Gamma_{mn}$. The vector $p$ can be expanded then effectively in the powers of the variation rate $\eta$:
\begin{equation}
p=\sum_{k=0}^{\infty} p^{(k)}\, ,
\label{e6a} \end{equation}
where $p^{(k)}\propto \eta^k$, and $p^{(0)}$ is the vector of the instantaneous equilibrium: $p^{(0)}_n=e^{-E_n(t)/k_B T}/Z$, $Z=\sum_n e^{-E_n(t)/k_B T}$. Substituting this expansion into the equation $\dot{p} = \gamma p$, using the fact that the instantaneous equilibrium satisfies the stationary master equation:
\begin{equation}
\gamma p^{(0)}=0\, ,
\label{e6b} \end{equation}
and that the time derivative adds one factor of the rate $\eta$, we see that the expansion (\ref{e6a}) reduces the master equation to the equations for the expansion terms: $\dot{p}^{(k)} = \gamma p^{(k+1)}$. Formally, these equations can be solved directly to produce the series of the recursion relations:
\begin{equation}
p^{(k+1)}=\gamma^{-1} \dot{p}^{(k)}\, .
\label{e6c} \end{equation}
However, since the matrix $\gamma$ of the transition rates has a non-vanishing kernel, spanned by $p^{(0)}$, its determinant is zero, and the calculation of the inverse matrix $\gamma^{-1}$ requires the introduction of a ''generalized'' inverse (also sometimes called ''pseudoinverse'') determined by some additional conditions. For the solution of the master equation, the appropriate generalized inverse is the ''group'' inverse (see, e.g., \cite{inv}) defined by the following relations:
\begin{equation}
\gamma \gamma^{-1} \gamma=\gamma \,  , \;\;\; \gamma^{-1}\gamma\gamma^{-1}=\gamma^{-1} \, , \;\;\; \gamma \gamma^{-1}=\gamma^{-1} \gamma\, .
\label{e7} \end{equation}
Combining the second and the third of these relations one sees, in particular, that the group inverse has the following important property:
\begin{equation}
\gamma^{-1}p^{(0)}=0\, .
\label{e7a} \end{equation}

Expression for the average of the element of heat $dQ$ (\ref{heat}), as it appears, e.g., in  Eq.~(\ref{first}), can be written in the matrix form using the vector-column $E=\{E_n\}$ of the energies $E_n$: $\langle dQ\rangle=-E^{\dagger} \gamma p dt= -E^{\dagger} \dot{p} dt$. The average heat transferred into the reservoir takes then the following form
\begin{equation}
\langle Q\rangle = \int \langle dQ\rangle = - \int E^{\dagger} \dot{p} dt  \, .
\label{e8} \end{equation}
We assume that the adiabatic evolution of the energies $E_n(t)$ starts and ends at times $t_i$ and $t_f$, with energies staying constant outside of this time interval. The time dependence of energies is also assumed to be sufficiently smooth, so that $\dot{E}_n(t)=0$ at both ends of the evolution, $t=t_i\, , t_f$. Substituting the expansion (\ref{e6a}) into Eq.~(\ref{e8})
one can see that the first, equilibrium, term $p^{(0)}$ of the expansion reproduces Eq.~(\ref{eq}) for the reversible heat transfer. To do this, we integrate Eq.~(\ref{e8}) by parts and express the equilibrium probabilities through the free energy $F=-T\ln Z$ of the system with the help of the relation $p^{(0)}_n=\partial F/ \partial E_n $ to obtain
\[ \langle Q\rangle^{(rev)} = -\int_{t_i}^{t_f} E^{\dagger} \dot{p}^{(0)} dt = \int_{t_i}^{t_f} \dot{E}^{\dagger} p^{(0)}dt  \]
\[-E^{\dagger} p^{(0)} \Big|_{t_i}^{t_f} =\int_{t_i}^{t_f} \sum_{n} \frac{\partial F}{ \partial E_n} dE_n - \sum_{n} E_n p^{(0)}_n  \Big|_{t_i}^{t_f} \]
\[ = \Delta F- \Delta U =- T \Delta S_{sys} \, . \]

The second term in the expansion (\ref{e6a}) gives the main contribution to the dissipated heat $\langle Q \rangle^{(dis)}$ in the adiabatic evolution:
\begin{eqnarray}
\langle Q\rangle^{(dis)} = -\int_{t_i}^{t_f} E^{\dagger} \dot{p}^{(1)} dt =
\int_{t_i}^{t_f} dt \dot{E}^{\dagger} p^{(1)} \nonumber \\
= \int_{t_i}^{t_f} dt \dot{E}^{\dagger} \gamma^{-1} \dot{p}^{(0)}\, ,
\label{e9} \end{eqnarray}
where we have used the fact that $p^{(1)}=\gamma^{-1} \dot{p}^{(0)}=0$ at $t=t_i\, , t_f$,  because $\dot{p}^{(0)}$ vanishes together with the derivatives $\dot{E}_n$ of the system energies at the ends of the evolution. Calculating the derivatives
\[\dot{p}^{(0)}_n= -\frac{\dot{E}_n}{k_B T}p^{(0)}_n - \frac{\dot{Z}}{Z}p^{(0)}_n \, ,\]
and making use of the property (\ref{e7a}) of the group inverse, we transform the expression for the heat dissipated irreversibly in the adiabatic evolution into the final form:
\begin{equation}
\langle Q\rangle^{(dis)} = -\frac{1}{k_B T} \int_{t_i}^{t_f} dt \sum_{n,m} \dot{E}_m[\gamma^{-1}]_{m,n} \dot{E}_n p^{(0)}_n\, .
\label{dis} \end{equation}

Equation (\ref{dis}) can be used for a system with arbitrary spectrum $E_n(t)$, as long as one can calculate the inverse $\gamma^{-1}$ of the transition rate matrix, as defined by Eq.~(\ref{e7}). As an example of the application of Eq.~(\ref{dis}), we calculate the dissipated heat for an adiabatic evolution of a two-state system with state energies $E_0(t)$ and $E_1(t)$. The calculation is simplified by the general property of Eq.~(\ref{dis}) which follows from Eq.~(\ref{e7a}): the dissipated heat depends, as should be, only on the energy differences in the energy spectrum $E_n(t)$, i.e., it does not change if all energies
are shifted by some time-dependent offset $\delta(t)$. For a two-state system, this means that the dissipated heat depends only on the time dependence of the energy difference $\epsilon (t)=E_1(t)-E_0(t)$ and not on the evolution of the individual energies $E_{0,1}$ separately. This means that, effectively, we can take the energies of the two states to be $0$ and
$\epsilon (t)$, and the Eq.~(\ref{dis}) is reduced as follows:
\[ \langle Q\rangle^{(dis)} = -\frac{1}{k_B T} \int_{t_i}^{t_f} dt \dot{\epsilon}(t)^2[\gamma^{-1}]_{11} p^{(0)}_1\, . \]

For a two-state system, transition matrix takes the following explicit form:
\[\gamma = \left( \!\!\! \begin{array}{cc} -\Gamma_{10} &  \;\; \Gamma_{01} \\ \Gamma_{10} \; & \!\!\! -\Gamma_{01} \!\!\!
\end{array} \right) , \]
and the group inverse defined by the conditions (\ref{e7}) can be calculated directly:
\begin{equation}
\gamma^{-1}=\gamma/\Gamma_{\Sigma}^2, \;\;\;\; \Gamma_{\Sigma} \equiv \Gamma_{10} +\Gamma_{01} \, .
\label{e10} \end{equation}
This result means that $[\gamma^{-1}]_{11}=-\Gamma_{01}/\Gamma_{\Sigma}^2$, and combining it with the detailed balance condition (\ref{e2}): $\Gamma_{10}/\Gamma_{01}=e^{-\epsilon/k_B T}$ and equilibrium probability $p^{(0)}_1=1/(e^{\epsilon/k_B T}+1)$ we obtain the heat dissipated in the adiabatic evolution of the two-state system with energy difference $\epsilon(t)$ between the states:
\begin{equation}
\langle Q\rangle^{(dis)} =\frac{1}{4k_B T} \int_{t_i}^{t_f} dt \frac{\dot{\epsilon}^2}{\Gamma_{\Sigma} \cosh^2(\epsilon/2k_B T)} \, .
\label{tls} \end{equation}

The higher-order terms in the expansion (\ref{e6a}) gives progressively smaller corrections to the heat $Q$ which vanish more rapidly with the rate $\eta$ of variation of the energies of the system. This means that in the adiabatic limit, one can take into account only the first two terms in the expansion Eq.~(\ref{e6a}), limiting the expression for the heat exchanged between
the system and the reservoir to the two contributions:
\begin{equation}
\langle Q\rangle= \langle Q\rangle^{(rev)} + \langle Q\rangle^{(dis)}
\label{tot} \end{equation}
which are given by Eqs.~(\ref{eq}) and (\ref{dis}), respectively, and correspond to the reversible and irreversible heat generated in the reservoir. One more useful remark that follow from Eq.~(\ref{tot}) is that the dissipated heat coincides with the ``dissipated work'' $\langle W\rangle^{(dis)}$, which can be defined naturally as the work done on the system in excess of the change of the system free energy. Indeed, combining Eqs.~(\ref{tot}) and (\ref{eq}) with the first law (\ref{first}) we see that
\begin{eqnarray}
\langle Q\rangle^{(dis)}=\langle Q\rangle-\langle Q\rangle^{(rev)} =
\langle W\rangle \nonumber \\
-\Delta U+T \Delta S_{sys} = \langle W\rangle -\Delta F  = \langle W\rangle^{(dis)},
\label{work} \end{eqnarray}
as stated above.

\subsection{Classical fluctuation-dissipation theorem}

Another feature of the heat dissipated in the adiabatic evolution that is important for operation of the reversible Maxwell's demon is the direct relation between the average dissipated heat given in general by Eq.~(\ref{dis}), or by Eq.~(\ref{tls}) in the case of a two-state system, and the magnitude of the  fluctuations of the heat exchanged between the reservoir and the system during the evolution. This relation, derived below, shows that in the adiabatic limit, the dissipated heat vanishes not only on average, but for every instance of the evolution implying that in principle, it can be made negligible in every cycle
of the Maxwell's demon operation. Qualitatively, the relation between the average dissipated heat and its fluctuations can be seen as a version of the classical limit of the fluctuation-dissipation theorem connecting the linear heat conductance $G_{th}$
and the zero-frequency spectral density $S_J (0)$ of the fluctuating heat flux in this conductance (see, e.g., \cite{LL}): $S_J (0)=2T^2G_{th}$. In the case of adiabatic evolution, this relation is replaced by the relation between the dissipated heat $\langle Q\rangle^{(dis)}$, which in this context is an analog of the thermal conductance, and the noise of this heat, which $\langle Q^2\rangle$ replaces the spectral density $S_J (0)$. This analogy stems from the fact that the dissipated heat in the adiabatic evolution is produced as a response to a small deviation from the equilibrium between the system and the reservoir, created by a finite rate of change of energies of the system in the same way as the average heat flow through the thermal conductance $G_{th}$ is produced by a small temperature bias across it.

To derive the relation between the average dissipated heat (\ref{dis}) and its noise quantitatively, we need to calculate the heat noise $\langle Q^2\rangle$ in the adiabatic evolution governed by the master equation (\ref{e1}). Taking the elementary heat (\ref{heat}) generated when the system is in the state $n$, and summing it over all states with their probabilities, we get the total elementary heat exchanged between the system and reservoir
\begin{equation}
dQ=\sum_{n,m}(E_n-E_m)\Gamma_{mn}\hat{p}_n dt\, .
\label{heat2} \end{equation}
In this expression, $\hat{p}_n$ denotes an arbitrary distribution of occupation probabilities which satisfies the rate equation (\ref{e1}), but does not necessarily coincide with the actual probability distribution in the evolution of the system because of, e.g., different initial condition imposed on the probabilities $\hat{p}_n$. Because of this, $dQ$ here is not the actual average heat as, e.g., in Eq.~(\ref{first}). Equation (\ref{heat2}) can be simplified further as follows:
\begin{eqnarray}
dQ=\sum_{n,m}(E_n-E_m)(\Gamma_{mn}\hat{p}_n-\Gamma_{nm}\hat{p}_m) dt/2 \nonumber \\ =\sum_{n,m}E_n(\Gamma_{mn}\hat{p}_n-\Gamma_{nm}\hat{p}_m)
dt=-\sum_n E_n \dot{\hat{p}}_n dt\, , \label{e11} \end{eqnarray}
where in the last step we used Eq.~(\ref{e1}), since the time dependence of the otherwise arbitrary probabilities $\hat{p}_n$ should still be governed by this equation. Using Eq.~(\ref{e11}) to define an expression for the heat $Q$ exchanged with the reservoir during the full evolution we get:
\begin{eqnarray}
Q=\int_{t_i}^{t_f} dQ =-\sum_n \int_{t_i}^{t_f} E_n \dot{\hat{p}}_n dt \nonumber \\
= \sum_n \Big[-E_n \hat{p}_n\Big|_{t_i}^{t_f} + \int_{t_i}^{t_f} \dot{E}_n \hat{p}_n dt \Big] \, . \label{e12} \end{eqnarray}

Equation (\ref{e12}) represents the heat generated in the reservoir as the difference between the work $W$ done on the system and the change of the system energy in the evolution process, $Q=W-(E_f-E_i)$. Since the system randomly occupies different energy levels in the evolution, the work $W$ is a fluctuating quantity similarly to the system energies $E_{f,i}$. They all contribute to the fluctuations of the generated heat defined as usual as
\[ \langle \tilde{Q}^2\rangle=\langle Q^2\rangle - \langle Q\rangle^2 . \]

Similarly to the average dissipated heat (\ref{dis}), we find these fluctuations in the first order in the small typical rate $\dot{E}_n$ of the variation of the system energies. Note that although Eq.~(\ref{dis}) contains the terms proportional to $\dot{E}_n^2$ under the integral over time, a reasonable adiabatic evolution changes the system energies considerably, i.e.
$\int_{t_i}^{t_f} dt \dot{E}_n \sim E_n$, so effectively, the dissipated heat (\ref{dis}) is of the first order in the evolution rate $\dot{E}_n$. In this approximation, the fluctuations of the initial and final energies of the system are given by their standard equilibrium expressions, $\langle \tilde{E}_{i,f}^2\rangle = C_{i,f}T^2$, where $C_{i,f}$ are the heat capacities of the system before and after the adiabatic evolution, as follows from the following considerations. First, corrections $p^{(1)}$ to the equilibrium probabilities
$p^{(0)}$, which could produce a non-equilibrium contributions to fluctuations of $E_{i,f}$ vanish at $t=t_i\, , t_f$ according to Eq.~(\ref{e6c}) under the adopted assumption of smooth evolution of the system energies. In addition, the correlations among different tunneling events described by the master equation decay on the time scale $\Gamma^{-1}$ set by the typical tunneling rate $\Gamma$, which is much shorter than the evolution time interval $t_f-t_i$. Because of this, the fluctuations of the initial and final energies of the system and the fluctuations of the work done during the adiabatic evolution are uncorrelated:
\begin{equation}
\langle \tilde{Q}^2\rangle =\langle \tilde{E}_i^2\rangle +\langle \tilde{E}_f^2\rangle +\langle \tilde{W}^2\rangle\, , \; W=\int_{t_i}^{t_f} \sum_n \dot{E}_n \hat{p}_n dt  \, .
\label{e13} \end{equation}
The fluctuations of the work $W$ done in the adiabatic evolution can be found according to the standard prescription for the correlation functions in the master-equation-governed evolution:
\begin{equation}
\langle \tilde{W}^2\rangle=\int_{t_i}^{t_f} dt dt' \sum_{n,m} \dot{E}_m(t') p_{m,n}(t',t) \dot{E}_n(t) p_n(t)   \, .
\label{e14} \end{equation}
Here $p_n(t)$ is the actual probabilities in the evolution process, and $p_{m,n}(t',t)$ is the distribution of probabilities evolving at time $t'$ out of the initial configuration at $t'=t$ in which the system occupies with certainty the state $n$: $p_{m,n}(t,t)=\delta_{m,n}$. Since Eq.~(\ref{e14}) already contains the rate of change of energies squared, the probabilities
in it can be calculated in the quasistatic approximation, in which $p_n(t)=p_n^{(0)}(t)$, and the master equation that determines the probabilities $p_{m,n}(t',t)$:  $\partial p/\partial t'=\gamma(t') p$, can  be simplified by essentially neglecting the time dependence of the transition rates, i.e. reducing it to the following form: $\partial p/\partial t'=\gamma(t)p$.
Formal solution of this equation can then be written down immediately:
\begin{equation}
p_{m,n}(t',t)= \big[e^{\gamma(t)|t'-t|}\big]_{m,n} \, .
\label{e15} \end{equation}
The fact that this expression is the same for $t'<t$ as for $t'>t$ is dictated by the fact that in the quasistationary approximation, the correlation functions should be symmetric with respect to the interchange of the two times, $t\leftrightarrow t'$. Combining Eq.~(\ref{e15}) with Eq.~(\ref{e14}), and taking into account that the other terms in the integrals evolve on the time scale much longer than that set by the tunneling rates, we can take the integral over $t'$ in Eq.~(\ref{e14}) by effectively integrating only the probabilities (\ref{e15}):
\begin{equation}
\langle \tilde{W}^2\rangle=-2 \int_{t_i}^{t_f} dt \sum_{n,m} \dot{E}_n \dot{E}_m p_n^{(0)} [\gamma^{-1}]_{m,n}  \, .
\label{e16} \end{equation}

Comparing Eq.~(\ref{e16}) to Eqs.~(\ref{dis}) and (\ref{work}) of the previous Subsection, we see that the fluctuations of the work done on the system are directly related to the dissipated part of the work:
\begin{equation}
\langle \tilde{W}^2\rangle=2 k_B T \langle W\rangle^{(dis)}  \, .
\label{e17} \end{equation}
The energy-fluctuation terms can be removed from Eq.~(\ref{e13}) for the heat noise by averaging over the initial and final states of the adiabatic evolution. Alternatively, in some situations, e.g. the one relevant for the most basic form of the reversible Maxwell's demon discussed below, the energy fluctuations become negligible automatically. In that case, fluctuations of the initial energy $E_i$ vanish because the evolution of the system starts from the definite (ground) state of the system separated by large energy gap $\Delta E \gg k_BT$ from the excited states, which suppresses all transitions that could change the system energy. The fluctuations of the final energy $E_f$ vanish because the evolution ends with the equally occupied two degenerate states of the system. Although the transitions between these two states are not suppressed, they do not cause any exchange of energy with the reservoir, since the energies of the two states of the system are the same. All other states are again separated from the two degenerate states by large energy gap suppressing the transitions to these states. For vanishing fluctuations of the energies $E_{i,f}$, Eq.~(\ref{e17}) can be extended directly to the relation for the noise of the heat transferred to the reservoir:
\begin{equation}
\langle \tilde{Q}^2\rangle=2 k_B T \langle Q\rangle^{(dis)}  \, ,
\label{fdt} \end{equation}
which can be viewed as an analog of the classical fluctuation-dissipation theorem for the thermal conductance in the situation of adiabatic evolution. The main consequence of this relation for the subsequent discussion of the reversible Maxwell's demon is the fact that the standard deviation of the heat noise vanishes together with the average dissipated heat, and as a result, in the adiabatic limit, the dissipated heat vanishes not only on average, but for each realization of the adiabatic evolution.

\section{Reversible Maxwell's demon}

The discussion of adiabatic evolution in the previous Section illustrates explicitly the standard understanding that thermodynamically optimum processes require sufficiently slow, adiabatic transformations, which minimize dissipated heat and work and make the whole process effectively reversible. Here we consider ``Maxwell's demon'' (for reviews, see \cite{MD}), a
device which extracts energy from thermal fluctuations in an equilibrium statistical system at temperature $T$ and transforms this thermal energy into free energy. All this is achieved by means of a process the central part of which is a measurement done on a fluctuating statistical system in thermal equilibrium, and application of the feedback control pulses which depend on the outcome of this measurement. In principle, Maxwell's demon (MD) can be realized using an arbitrary quantum system with the energy spectrum $E_n$ weakly interacting with a thermal reservoir, as considered in the previous Section. The simplest cycle that realizes the demon
operation consists of three steps.

The first step is an adiabatic evolution of the system extracting heat from a thermal reservoir in equilibrium. For simplicity, we assume that the system has a configuration of the energy levels $E_n$ such that the ground state energy $E_0$ is well separated from the excited states, so that there is a range of temperatures $T$ in which the system occupies the ground
state with near certainty, so that the entropy (\ref{e3}) of the system effectively vanishes. The adiabatic evolution should start then with the system in the unique ground state and should  bring it to another equilibrium configuration with the level occupation probabilities $p_n$, in which the entropy (\ref{e3}) is increased to a nonvanishing value $S_{sys}$. According to Eq.~(\ref{eq}), in the course of this evolution, the heat
\begin{equation}
\langle Q \rangle = -k_B T\sum_{n} p_n \ln p_n
\label{e18} \end{equation}
is extracted on average from the reservoir. As follows from the discussion in the previous Section, in the adiabatic limit, the fluctuations of the extracted heat in this case are determined by the fluctuations of the system energy at the final point of the evolution. If there are no energy fluctuations at the final point, the fluctuations of $Q$ vanish and
extracted heat $Q$ is given by Eq.~(\ref{e18}) for each individual cycle, $Q=\langle Q \rangle$. The adiabatic evolution considered here is reversible and generates entropy $S_{sys}$
in the system's states. If the system represents a computing device, this entropy corresponds to the information stored in this device. The reversal of the adiabatic evolution just described would remove this information from the device by bringing it to the definite prescribed state, and therefore coincides with the Landauer information erasure discussed in the Introduction. Thus, the first step of the operation of our Maxwell's demon is the reversal
of the information erasure protocol.

The second step of the MD operation is the measurement of the actual state of the system at the end of the evolution. Since we are limiting the discussion here to classical situations, the only two important features of the measurement is whether it is precise, and whether it is done reversibly, without the avoidable energy dissipation. If the measurement is ideal in
both respects, then the state of the system at the end of the evolution is determined precisely, and the only thermodynamic price paid for this is the need to physically record this information $I$ about the state of the system. For an ideal measurement, the statistics of the measurement outcomes exactly repeats the statistics of the occupation of the energy levels of the system, and is characterized by the probabilities $p_n$. This means that in the limit of large number of MD operation cycles, each measurement generates on average the amount of information
\begin{equation}
I=-\sum_{n} p_n \ln p_n
\label{e19} \end{equation}
which coincides with the system entropy before the measurement. This fact illustrates the general understanding of information as the entropy of the computing system that was discussed in the Introduction.

The last step of the MD cycle is the application of the sequence of pulses to the system that drive it quickly from the actual observed state at the end of the adiabatic evolution into the
configuration when the occupied state is the unique ground state of the system, similar to the starting point of the evolution. The pulses should be applied in a time much shorter than the typical transition rates. Since the energy spectrum of the system is controlled by the external pulses, the ground state can in general be a different state in each cycle (depending on the outcome of the measurement) but can be made to have the same energy each time. Such a rapid return to the ground states completes the MD cycle. It does not generate any heat in the reservoir, since no transitions have time to happen during it. The net result of one cycle of MD operation is then the extraction of the heat (\ref{e18}) from the reservoir, which, by the first law, was transformed into the free energy of the external source manipulating the energy states of the system. This transformation was achieved at a cost of creation of the information (\ref{e19}) about the actual state of the fluctuating system at the end of the adiabatic evolution. These expressions shows that the operation of such a optimal reversible demon does not contradict the second law. Reversible erasure of the information (\ref{e19}) dissipates back into heat precisely the energy $k_BTI$ extracted by the demon from the thermal reservoir.

\begin{figure}[t]
\centering
\includegraphics[bb=-5 0 380 145, clip=true, scale=0.63]{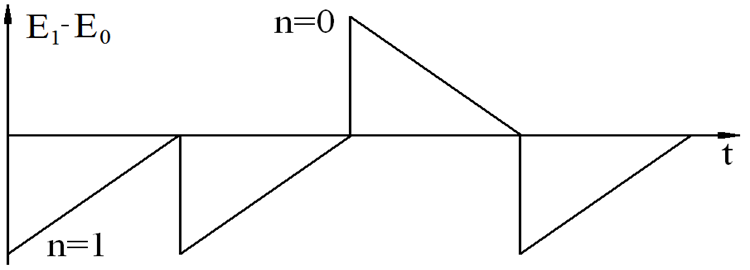}
\caption{Typical qualitative time dependence of the energy difference between the two states of an effectively two-state system, e.g. single-electron box, that corresponds to several cycles of the operation of reversible Maxwell's demon. Two parts of each cycle that can be explicitly seen in the diagram are: (1) adiabatic ramp of the energy difference starting from initial value that is much larger in magnitude than the thermal energy $k_BT$ and ending at the degeneracy point where the energy difference vanishes; and (2) abrupt return from the degeneracy point to the ground state. At the beginning of the ramp, the system is with certainty in the ground state, while at the degeneracy point both states are occupied with equal probability 1/2. Whether the applied feedback makes the state $n=0$ or $n=1$ the ground state in a given cycle depends on the actual state the measurement finds the system in at the degeneracy point in this cycle. The heat $k_BT \ln 2$ is extracted from the reservoir during each adiabatic ramp and converted into the free energy transferred to the external source which manipulates the energies of the system.}
\label{MD-op} \end{figure}

The demon operation just described relies on the existence of the accurate measurements of the states of the system that are distinguished by their energies $E_n(t)$. While theoretically the energy is an absolutely legitimate observable, experimentally, the non-invasive measurements of energy presents  considerable problems. In this respect, practical implementation of the MD cycle is only possible when the energy states $n$ of the system can also be distinguished by some other physical quantity that is directly measurable. This is the case for the single-electron circuits, where the dominant energy of the system is the electrostatic energy, and the different energy states are distinguished also by the excess number of individual electron charges on the electrodes of the system. The charge states can be measured directly with any of the several existing single-charge detectors, DC \cite{al2,fulton,saira2} or RF \cite{rfset} single-electron transistors, or quantum-point contacts \cite{qpc}. Such a Maxwell's demon operating with individual electrons has been realized \cite{we1} using the most basic single-electron system, single-electron box (SEB) \cite{mb,laf}: two electrodes connected by a small tunnel junction that allows electrons to tunnel between them. An advantage of this system is that at low temperatures, $k_B T\ll E_C$, where $E_C$ is the characteristic electrostatic energy associated with one electron charging the system, its dynamics can be naturally reduced to that of a two-state system, simplifying the measurement and control part of the MD cycle. Moreover, the energies $E_{0,1}$ of the two remaining states of the two-state system can be gate-voltage driven back-and-forth between the regimes of large $|E_0-E_1|\gg k_BT$, and small, $|E_0-E_1|\ll k_BT$, energy separation, as needed for the MD operation - see Fig.~\ref{MD-op} and the discussion below.

In the two-state regime, only the two charge states, $n=0$ and $n=1$, have the non-vanishing occupation probabilities. These states differ by one extra electron charge residing either on the left or on the right electrode of the SEB. In this respect, the SEB Maxwell's demon is similar to the Szilard engine \cite{szilard}, in which the states of the Maxwell's demon working substance are distinguished by two possible positions of one particle. This correspondence between the energy and charge states also makes it possible to measure directly the actual state of the system with a charge detector. The energies $E_0$ and $E_1$ of the two charge states can be controlled by the applied time-dependent gate voltage (for more detailed  description of the system -- see, e.g., \cite{we2}. As partly discussed in the previous Section, the thermodynamic properties of a two-state system, including now the characteristics of the MD cycle, depend only on the energy difference $\epsilon =E_1(t)-E_0(t)$ and not on the individual energies $E_{0,1}$ separately. The gate-voltage-driven typical qualitative time dependence of the energy difference that corresponds to several cycles of MD operation as described above is shown in Fig.~\ref{MD-op}.

Each cycle starts with an adiabatic ramp that slowly drives the system from the state with $|\epsilon|\gg k_BT$, when the ground state is occupied with near absolute certainty, to the degeneracy point $\epsilon =0$, where the two states of the system are equally occupied. This means that the entropy of the system is increased by $\Delta S_{sys}=k_B\ln 2$ and the heat $Q= k_BT \ln 2$, as in Eq.~(\ref{eq}), is extracted from the reservoir during each ramp. Note that under the
conditions of this cycle, there are no equilibrium fluctuations of the heat at the end point of the ramp, and therefore, no heat fluctuations at all in the adiabatic limit, i.e., the extracted heat is the same in each cycle. The cycle of MD operation is completed by the measurement of the actual state of the system at degeneracy point, and application of the corresponding feedback pulse of gate voltage which swiftly turns the measured state into the ground state of the system. Since the pulse is fast on the time scale set by the transition rates, no heat is exchanged with the reservoir in this return to the ground state, so the net result of the cycle is the extraction of the heat during the adiabatic ramp, which is converted into the free energy transferred to the source of the control pulses. For instance, in the case of SEB, this means that the battery producing the gate voltage, ideally, should be charged by the MD operation. Thermodynamic cost of this transformation of heat into free energy is creation of one bit of information about the actual state of the system at degeneracy; for the SEB,
position of the extra electron on the left or right electrode of the box. If this information is erased in an optimal, reversible, fashion, precisely the same amount of free energy as extracted by the Maxwell's demon is converted back into heat, ensuring that the second law of thermodynamics is satisfied.

\subsection{Minimization of the dissipated heat}

Ideal operation of the Maxwell's demon discussed so far is affected in general by several possible imperfection. The most fundamental limitation is imposed by the finite rate of the adiabatic ramp which, in addition to the reversible extraction of the heat (\ref{eq}) from the reservoir, dissipates irreversibly the heat (\ref{dis}). In the case of the MD cycle based on a two-state system, e.g. the SEB described above, the irreversibly dissipated heat is given by Eq.~(\ref{tls}) and can be minimized by appropriate choice of the profile of the energy difference ramp $\epsilon(t)$. To do this, one needs to minimize the integral in Eq.~(\ref{tls}) with respect to the function $\epsilon(t)$. Since the integral in Eq.~(\ref{tls}) does not contain any derivatives of $\epsilon(t)$ of higher than the first order, this minimization problem is equivalent to the one encountered, e.g., in the Lagrangian formulation of classical mechanics, and leads to the Lagrange equation for the function $\epsilon(t)$ minimizing the dissipated heat:
\begin{equation}
\frac{d}{dt} \frac{\partial J}{\partial \dot{\epsilon}} =\frac{\partial J}{\partial \epsilon} \, , \;\; J\equiv \frac{\dot{\epsilon}^2}{4k_B T \Gamma_{\Sigma} (\epsilon)\cosh^2(\epsilon/2k_B T)} \, .
\label{min1} \end{equation}
Calculating the derivatives, one can see that the Lagrange equation can be transformed into the form requiring that the integrand in Eq.~(\ref{tls}) is constant in time:
\begin{equation}
\frac{dJ}{dt} =0 \, ,\; \mbox{i.e.} \; \frac{\dot{\epsilon}}{\sqrt{\Gamma_{\Sigma}(\epsilon)} \cosh(\epsilon/2k_BT)} =-\sqrt{4k_BTJ}\, .
\label{min2} \end{equation}
Integrating this equation over the time interval of adiabatic evolution, from $t_i=0$, when the energy difference acquires some large value $\epsilon(0)= \epsilon_i\gg T$, to $t_f=\tau$, when $\epsilon(\tau)=0$, one obtains the condition that gives the
constant heat flux $J$ during the evolution:
\begin{equation}
\sqrt{4k_B TJ} \tau= \int_0^{\epsilon_i}\frac{d\epsilon}{\sqrt{\Gamma_{\Sigma}(\epsilon)} \cosh( \epsilon/2k_B T)} \, ,
\label{min3} \end{equation}
and therefore, determines the total dissipative heat $\langle Q\rangle^{(dis)}$ (\ref{tls}) in the adiabatic ramp:
\begin{equation}
\langle Q\rangle^{(dis)} =J\tau  \, .
\label{min4} \end{equation}

To obtain the explicit solution of Eq.~(\ref{min3}), one needs to specify the energy dependence of the total tunneling rate $\Gamma_{\Sigma}(\epsilon)$ in the two-state system. An important example is provided by the hybrid SEB based on the normal metal/insulator/superconductor (NIS) tunnel junction as used in the experiment (\cite{we1}). Single-electron tunneling rate in the NIS junction is (see, e.g., \cite{we2})
\begin{equation}
\Gamma_{\Sigma}(\epsilon) = 2 \Gamma_m \cosh^2 \{ \epsilon/2k_BT\} \, ,
\label{min5} \end{equation}
where $\Gamma_m$ is the tunneling rate at degeneracy, when $\epsilon=0$,
\[ \Gamma_m =\frac{1}{R_Te^2} (2\pi \Delta k_B T)^{1/2} e^{-\Delta /k_BT} \, . \]
Here $R_T$ is the normal-state resistance of the NIS tunnel junction and $\Delta$ is the superconducting energy gap in the superconducting electrode of the junction. With the rate (\ref{min5}), integral in Eq.~(\ref{min3}) can be calculated explicitly to find the minimum $Q_m$ of the dissipated heat $\langle Q\rangle^{(dis)}$ (\ref{tls}) for the optimum adiabatic ramp satisfying the condition (\ref{min2}):
\begin{equation}
Q_m=J\tau=\frac{k_BT}{2\tau \Gamma_m} \tanh^2\frac{\epsilon_i}{2k_BT}\, .
\label{min6} \end{equation}
This equation shows that, qualitatively, the dissipated heat is indeed minimized for slow adiabatic ramps, when the total ramp time $\tau$ is large, $\tau\gg \Gamma_m^{-1}$, and that in this limit, the dissipated heat can in principle be made negligible in comparison with the reversible heat $k_BT \ln 2$ extracted from the reservoir in the ideal MD cycle.

It is instructive to compare Eq.~(\ref{min6}) to the result that would be obtained without the minimization of the dissipated heat, e.g., for the linear time dependence of the energy difference $\epsilon(t)$ in the adiabatic evolution shown qualitatively in Fig.~\ref{MD-op}. Evaluating the integral in Eq.~(\ref{tls}) for $\dot{\epsilon}=\mbox{const}=\epsilon_i/\tau$ and the total tunneling rate (\ref{min5}), one obtains:
\begin{equation}
\langle Q\rangle^{(dis)}=\frac{\epsilon_i}{4\tau \Gamma_m} \tanh\frac{\epsilon_i}{2k_BT}
\big[1-\frac{1}{3}\tanh^2\frac{\epsilon_i}{2k_BT}\big] \, .
\label{min66} \end{equation}
We see that in the relevant limit of large initial energy $\epsilon_i\gg k_BT$, non-optimized dissipative heat (\ref{min66}) is larger than the optimized heat (\ref{min6}) by a large factor $\epsilon_i/3k_BT$. This shows that the optimization procedure can play quite an important role in the MD operation.

We can also find explicitly the time dependence of the energy difference ramp $\epsilon(t)$ for which the dissipated heat $\langle Q\rangle^{(dis)}$ reaches the minimum (\ref{min6}). Integrating Eq.~(\ref{min2}) with the NIS tunneling rate (\ref{min5}), we obtain
\[ t/\tau = 1-\tanh(\epsilon(t)/2T)/\tanh(\epsilon_i/2k_BT) \, . \]
Finally, solving this equation for $\epsilon(t)$, one finds the profile of the optimum adiabatic ramp that minimizes the dissipated heat in the hybrid SEB:
\begin{equation}
\epsilon(t)=k_B T\ln \Big[\frac{2\tau-(1-e^{-\epsilon_i/k_BT})t}{2\tau e^{-\epsilon_i/k_BT}+(1-e^{-\epsilon_i/k_BT})t} \Big]\, .
\label{min7} \end{equation}
In this case, as one can see from Eqs.~(\ref{min2}) and (\ref{min5}), $|\dot{\epsilon}|\propto \cosh^2 \{ \epsilon/2k_BT\}$, i.e., qualitatively, Eq.~(\ref{min7}) describes more rapid variation of the energy difference $\epsilon (t)$ at large $\epsilon$, where the total tunneling rate is larger, and slower variation in the vicinity of the degeneracy point.

\section{Conclusions}

In this work, we have discussed the reversal of the optimal Landauer's information erasure. This reversal serves as the central element of the Maxwell's demon operating at the limit of thermodynamic efficiency. The limit is consistent with the second law of thermodynamics and corresponds to extraction of $k_BT\ln 2$ of energy from an equilibrium thermal reservoir at temperature $T$ per one bit of generated information about the working substance of the demon. The reversible Maxwell's demon considered here can be implemented not only with individual electrons in single-charge structures as in \cite{we1}, but also in other nanostructures, e.g. molecular systems, or in the Josephson junction structures based on the dynamics of individual magnetic flux quanta.

\begin{acknowledgement}
This work was supported in part by the U.S. NSF under the grant PHY-1314758 (D.V.A.) and by Academy of Finland under the grant no.\ 272218 (J.P.P.).
\end{acknowledgement}


\begin{thebibliography}{[1]}

\bibitem{we1} J.\,V. Koski, V.\,F. Maisi, J.\,P. Pekola, and D.\,V. Averin, Proc.\ Nat.\ Acad.\ Sci. \textbf{111}, 13786 (2014).

\bibitem{bustamante} C. Bustamante, J. Liphardt, and F. Ritort, Phys.\ Today \textbf{58}, 43 (2005).

\bibitem{campisi} M. Campisi, P. H\"{a}nggi, and P. Talkner, Rev.\ Mod.\ Phys. \textbf{83}, 771 (2011).

\bibitem{seifert} U. Seifert, Rep.\ Prog.\ Phys. \textbf{75}, 126001 (2012).

\bibitem{pekola} J.\,P. Pekola, Nat.\ Phys. \textbf{11}, 118 (2015).

\bibitem{millen} J. Millen and A. Xuereb, New J.\ Phys. \textbf{18}, 011002 (2016).

\bibitem{toyabe} S. Toyabe, T. Sagawa, M. Ueda, E. Muneyuki, and M. Sano, Nat.\ Phys. \textbf{6}, 988 (2010).

\bibitem{koski} J.\,V. Koski, A. Kutvonen, I.\,M. Khaymovich, T. Ala-Nissila, and J.\,P. Pekola, Phys.\ Rev.\ Lett. \textbf{115}, 260602 (2015).

\bibitem{QMD} J.\,P. Pekola, D.\,S. Golubev, and D.\,V. Averin, Phys.\ Rev.\ B \textbf{93}, 024501 (2016).

\bibitem{pt} E. Lutz and S. Ciliberto, Phys.\ Today \textbf{68}, Issue 9, p.\ 30 (2015); D.\,V. Averin, ibid.\ \textbf{69}, Issue 8, p.\ 12 (2016).

\bibitem{rev} J. Ren, V.\,K. Semenov, Yu.\,A. Polyakov, D.\,V. Averin, and J.\,S. Tsai, IEEE Trans.\ Appl.\ Supecond. \textbf{19}, 961 (2009).

\bibitem{landauer} R. Landauer, IBM J.\ Res.\ Devel. \textbf{3}, 183 (1961); Nature \textbf{335}, 779 (1988).

\bibitem{exp1} A. B\'erut, A. Arakelyan, A. Petrosyan, S. Ciliberto, R. Dillenschneider, and E. Lutz, Nature \textbf{483}, 187 (2012).

\bibitem{exp2} Y. Jun, M. Gavrilov, and J. Bechhoefer, Phys.\ Rev.\ Lett. \textbf{113}, 190601  (2014).

\bibitem{exp3} J.\,P.\,S. Peterson, R.\,S. Sarthour, A.\,M. Souza, I.\,S. Oliveira, J. Goold, K. Modi, D.\,O. Soares-Pinto, and L.\,C. C\'eleri, Proc.\ R.\ Soc.\ A \textbf{472}, 20150813 (2016).

\bibitem{bennett} C.H. Bennett, Studies in Hist.\ and Phil.\ of Mod.\ Phys. \textbf{34}, 501 (2003).

\bibitem{brillouin} L. Brillouin, Science and Information Theory, (Acad. Press, 1960), Ch.\ 13.

\bibitem{al1} D.\,V. Averin and K.\,K. Likharev, in: Mesoscopic Phenomena in Solids, ed. by B. Al'tshuler, P. Lee and R. Webb (Elsevier, Amsterdam, 1991), p.\ 173.

\bibitem{we3} D.\,V. Averin and J.\,P. Pekola, Europhys.\ Lett. \textbf{96}, 67004 (2011).

\bibitem{kung} B. K\"ung, C. R\"ossler, M. Beck, M. Marthaler, D. S. Golubev, Y. Utsumi, T. Ihn, and K. Ensslin, Phys.\ Rev. X \textbf{2}, 011001 (2012).

\bibitem{saira} O.-P. Saira, Y. Yoon, T. Tanttu, M. M\"ott\"onen, D.\,V. Averin and J.\,P. Pekola, Phys.\ Rev.\ Lett. \textbf{109}, 180601 (2012).

\bibitem{irrev} M. L\'opez-Su\'arez, I. Neri, and L. Gammaitoni, Nat.\ Comm. \textbf{7}, 12068 (2016).

\bibitem{strasberg} Ph. Strasberg, G. Schaller, T. Brandes, and M. Esposito, Phys.\ Rev.\ Lett. \textbf{110}, 040601 (2013); Phys.\ Rev.\ E \textbf{88}, 062107 (2013).

\bibitem{mb} M. B\"uttiker, Phys.\ Rev. B, \textbf{36}, 3548 (1987).

\bibitem{laf} P. Lafarge, H. Pothier, E.\,R. Williams, D. Esteve, C. Urbina, and M.\,H. Devoret, Z.\ Phys. B \textbf{85}, 327 (1991).

\bibitem{inv} S.\,L. Campbell and C.\,D. Meyer, Generalized inverses of linear transformations
(SIAM, Philadelphia, 2009).

\bibitem{LL} E.\,M. Lifshitz and L.\,P. Pitaevskii, Statistical Physics, Part II (Pergamon, Oxford, 1980), Sec.~88.

\bibitem{MD} H.\,S. Leff and A.\,F. Rex (eds.), Maxwell's Demon 2 (IOP Publishing, 2003).

\bibitem{we2} J.\,P. Pekola, O.-P. Saira, V.\,F. Maisi, A. Kemppinen, M. M\"ott\"onen, Yu.\,A. Pashkin, and D.\,V. Averin, Rev.\ Mod.\ Phys. \textbf{85}, 1421 (2013).

\bibitem{al2} D.\,V. Averin and K.\,K. Likharev, J.\ Low Temp.\ Phys.  \textbf{62}, 345 (1986).

\bibitem{fulton} T.\,A. Fulton and G.\,J. Dolan, Phys.\ Rev.\ Lett. \textbf{59}, 109 (1987).

\bibitem{saira2} O.-P. Saira, M. M\"ott\"onen, V.\,F. Maisi, and J.\,P. Pekola, Phys.\ Rev.\ B \textbf{82}, 155443 (2010).

\bibitem{rfset} R.\,J. Schoelkopf, P. Wahlgren, A.\,A. Kozhevnikov, P. Delsing, D.\,E. Prober, Science \textbf{280}, 1238 (1998).

\bibitem{qpc} M. Field, C.\,G. Smith, M. Pepper, D.\,A. Ritchie, J.\,E.\,F. Frost, G.\,A.\,C. Jones, and D.\,G. Hasko, Phys.\ Rev.\ Lett. \textbf{70}, 1311 (1993).

\bibitem{szilard} L. Szilard, Z.\ Phys.  \textbf{53}, 840  (1929).

\end{thebibliography}
\end{document}